\documentclass[submission,copyright,creativecommons]{eptcs}
\providecommand{\event}{Wivace 2013} 
\usepackage{breakurl}             
\usepackage{graphicx}

\title{Semi-structured data extraction and modelling: the WIA Project}
\author{Gianluca Colombo, Ettore Colombo, Andrea Bonomi
\institute{Technological Transfer Consortium - C2T, Milan, Italy}
\email{giacolos@gmail.com}
\and
Alessandro Mosca
\institute{KRDB Research Centre\\
Free University of Bozen-Bolzano, Italy}
\email{mosca@inf.unibz.it}
\and
Simone Bassis
\institute{Department of Computer Science\\
University of Milano, Italy}
\email{bassis@di.unimi.it}
}
\def\titlerunning{Semi-structured data extraction and modelling: the WIA Project}
\def\authorrunning{G. Colombo, E. Colombo, A. Bonomi, A. Mosca \& S. Bassis}
\begin{document}
\maketitle



Over the last decades, the amount of data of all kinds available electronically has increased dramatically. Data are accessible through a range of interfaces including Web browsers, database query languages, application-specific interfaces, built on top of a number of different data exchange formats. The management and the treatment of this large amount of data is one of the biggest challenges for Information Technology these days \cite{Palm012, Fabian013, Nitin013, Ger012}. 
All these data span from un-structured (e.g. textual as well as multimedia documents, and presentations stored in our PCs) to highly structured data (e.g. data stored in relational database systems).  Very often, some of them have structure even if the structure is implicit, and not as rigid or regular as that found in standard database systems (i.e., not table-oriented as in a relational model or sorted-graph as in object databases). Following \cite{Abiteboul1997}, we refer here to this kind of data in terms of ``semi--structured'', which sometimes are also called ``self-describing''. Their main characteristics are: no fixed schema, the structure is implicit and irregular, and they are nested and heterogeneous.

Spreadsheet documents perfectly adhere to the above definition of semi-structured data.
Besides the freedom and flexibility that they offer to build and manipulate data, the pervasive diffusion of end-users software development tools \cite{Ballinger:2003:SVI:857080.857093} is among the main reasons why semi-structured data have tremendously grown over the last years, above all spreadsheets. The planetary use of spreadsheets is widely witnessed by many studies spanning from software engineering to business management \cite{DBLP:journals/corr/abs-0802-3924}, which unanimously identify the spreadsheets flexibility as the key factor for their success. Spreadsheets are the lightweight technology able to supply companies with easy to build business management and business intelligence applications, and business people largely adopt spreadsheets as smart vehicles for data files generation and sharing. Moreover, analysts \cite{Auto2009, Inter2008, over2008, SpreadEng2007} agree on the fact that spreadsheet continue to co-exist even though more SMEs upgrade their business applications for reasons of planning, budgeting, and control: nothing seems to be as tailored to the particular SME businesses as the spreadsheet technologies. This explains why enterprises of every kind produce large amounts of spreadsheets, and spreadsheets are almost everywhere around us. 

The algorithmic approach to the problem of automatic data structure extraction from grid-structured and free topological-related data hereby described precisely emerges from the WIA project, i.e. a research project in the field of computer--based tools to support business activities, carried out by the Technological Transfer Consortium (C2T)\footnote{http://consorzioc2t.it/}  on behalf of Rho Inform srl\footnote{Rho Inform is a small sized enterprise whose core business is on balance date mining http://www.rhoinform.com/ .} . The attention was on the use of electronic worksheets (i.e. Excel), their role in business process and the value this artifacts intrinsically held with special attention to Small and Medium Enterprise (SMEs). WIA stands in fact for Worksheets Intelligent Analyzer. Thus, the grounding question behind WIA is: how to enrich spreadsheet tools with a set of functionalities based on the visual interaction for analyzing, design, modifying and using worksheets in order to improve the understanding and the exploitation of worksheet--based artifacts \cite{cabitza2013leveraging}? As a consequence to this question, the development phase of WIA concerned the building of a proof of concepts for a tool able to extract useful information from Excel Worksheets, represent them in a graph--based visualization and managing their contents by manipulating the graph--based visualization. The WIA prototype by now works on Excel spreadsheets but the conceptual architecture under this implementation can be applied also to other spreadsheet tools, such as Google Docs and LibreOffice.

\begin{figure}
\centering
  \includegraphics[width=12cm]{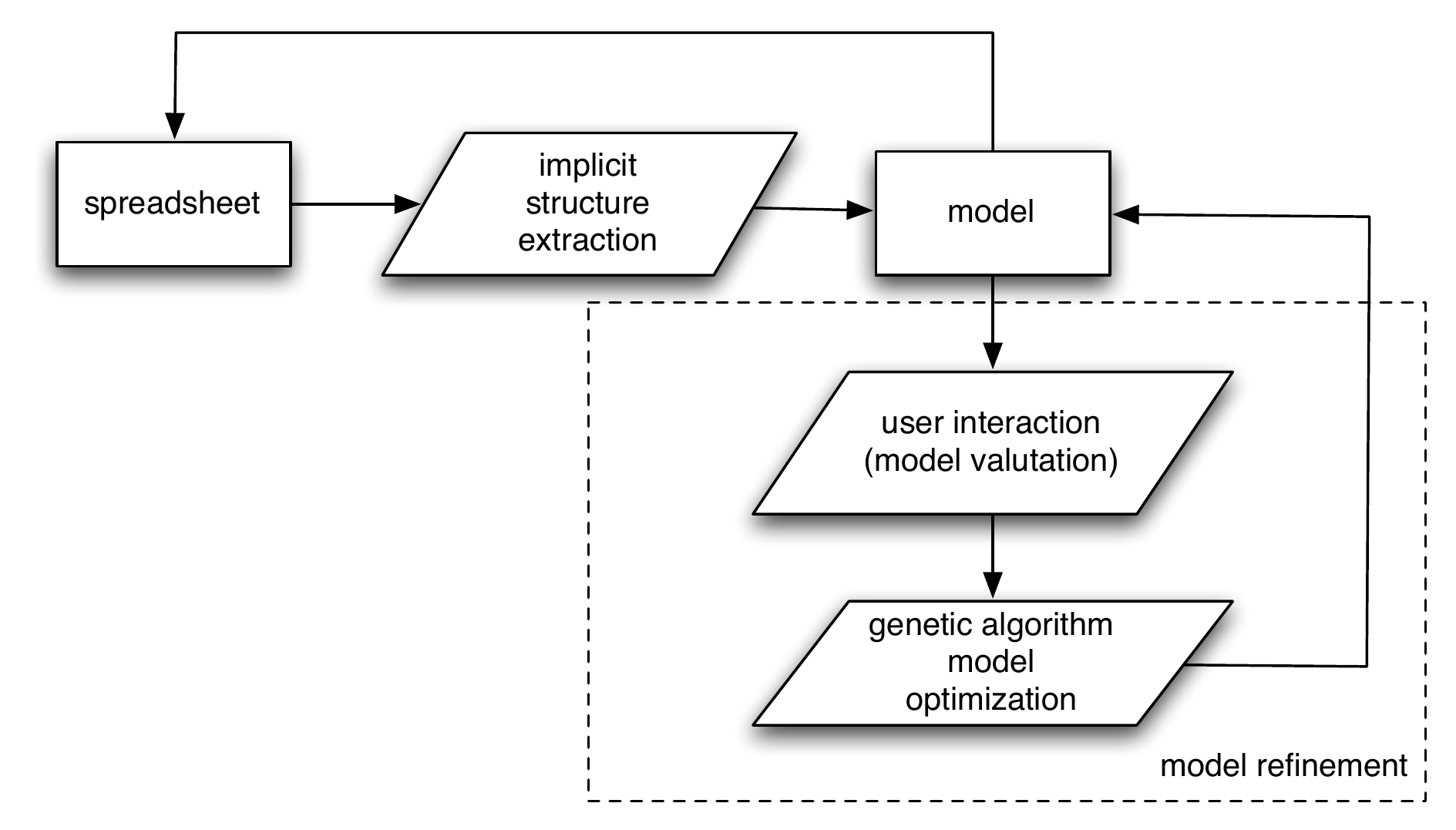}
   \caption{High level workflow.}
   \label{fig:workflow}
\end{figure}

The absence of an explicit schema definition in the spreadsheet design has a twofold impact on the use of this technology: on the one hand, it provides the flexibility the users are looking for and makes it easier for non-expert programmers to immediately start playing an active role in the enterprise business activities, on the other, it provides a degree of freedom that is often prone to erroneous spreadsheet creation. Moreover, the lack of any conceptual layer offering a higher-level abstraction over the co-presence of data and operations in a spreadsheet, makes the understanding of the overall spreadsheet functionalities an extremely difficult, and often ambiguous, process: the functionalities might be partially, or completely, misunderstood by the current user of the spreadsheet, thus giving rise to spreadsheet updates that are in contrast with the intended semantics of original arrangement of data and operations in the grid (as one would expect, the problem gets further exacerbated when a spreadsheet is intended to be used or modified in a collaborative working environment) .  


The WIA-algorithm hereby briefly introduced shows how to provide a description of spreadsheet contents in terms of higher level of abstractions or conceptualizations \cite{Floridi:2008:MLA:1404308.1404338}. The WIA-algorithm target is about the extraction of i) the calculus work-flow implemented in the spreadsheets formulas and ii) the logical role played by   the data which take part into the calculus. The aim of the resulting conceptualizations is to provide  spreadsheets with abstract representations useful for further model refinements and optimizations through evolutionary algorithms computations.
From the WIA-algorithm standpoint, the spreadsheet is described in terms of  cells, groups of cells and formulas. 
According to the fact that instantiated cells can be gathered by considering i) the roles they play in specific formulas and ii) the position they occupy in the spreadsheet, two main types of cells groups are defined: formula--based and topology--based groups. 
Finding formula--based groups is quite trivial: cycling over the entire set of spreadsheet formulas, for each instantiated cell the whole set of instantiated cells involved in the calculus of its value (i.e. the result of the partial calculus)are parsed by WIA-algorithm. As a result, a direct graph representation of the spreadsheet data--flow is provided.
Nevertheless, such a representation gives little information about the  spreadsheet data conceptual organization. Very often topological choices made by the spreadsheet developer work as implicit vehicles for data conceptualization \cite{cabitza2013leveraging} as in  the case of tables or the use of different sheets within the same workbook.

In order to make the algorithm able to extract this hidden data--structure, the basic concept  of ``segment'' has been defined. A segment is a list of contiguous cells respecting the following rules: 

    \begin{description}
    \item[R1] Cell contiguity over the scan direction, i.e cells must be subsequent;
    \item[R2] Data type continuity, i.e. cells must have the same data type;
    \item[R3] Formula structure continuity, i.e. cells must have the same formula string when the operands are stripped off, that is they must share the same normalized formula;
    \item[R4] Reference continuity, i.e formula operands in contiguous cells must point to contiguous cells in any referred segment;
    \item [R5] Non-intersection, i.e. no cell can belong to more than one group having cardinality greater than one.
    \end{description}

Working on segments, the WIA-algorithm follows these steps:

 \begin{description}
    \item[1 - BuildMatrix]     In this step the algorithm builds a 2D matrix representation of the used cells.
    \item[2 - GenerateSegments(ScanDirection.Vertical)]
    \item[3 - GenerateSegments(ScanDirection.Horizontal)]      In these steps the matrix is sequentially scanned over horizontal and vertical direction, building segment candidates that satisfy the given requirements R1, R2 and R3.
    \item [4 - ParseFormulaSegments] In this step, the algorithm cycles over all candidate segments and finds reference relations between them. The goal of this step is to check R4 requirement on Segment candidates.
    \item[5 - GetPreferredScanDirection] In this step  vertical or horizontal direction is chosen according to a simple heuristic.
     \item[6 - PruneSegments(preferredScanDir] In this step performs a pruning of segments such as the non-intersection of segments requirement R5 is met.
     \item[7 - SetReferencedSegments] This step finalizes Segment objects, populating the reference list to other cells.
      \item[8 - GenerateGroups] Finally, each selected Segment is converted into a group in order to be merged with the formula--based groups.
      \end{description}
       

Despite the undeniable fact that data and operations over these data (e.g. references,  groups, or operations) are considered as first--class citizens in spreadsheets, and real enterprise applications are built on top of their integrated arrangement, there is at least a third silently operating actor in the way the spreadsheet enters the business activities. Each spreadsheet is defined (and possibly maintained) according to an implicit representation of the domain knowledge, often called ``conceptual layer'' (``Spreadsheets are the result of a mapping of higher-level abstract models in the user mind to a simple two-dimensional grid structure"~\cite{Erwig:2005:AGM:1062455.1062494}). In other terms, along the ``spreadsheet paradigm" of programming and building enterprise applications, the domain knowledge driving the underlying infrastructure of the programs and applications, even though present, is completely hidden into the topological disposition of the cells in the two-dimensional grid space, the layouting, and the internal dependencies expressed by the operations themselves.

Actually, the more spreadsheets grow in complexity (e.g., their use in product development plans and quoting), the more their arrangement, maintenance, and analysis appear as a knowledge-driven activity. As a matter of fact, the actual spreadsheet development environments do not supply any conceptual modeling need as required by the use of complex spreadsheets, thus leaving enterprises business management needs uncovered. Current spreadsheet editors supply users with services to handle the sheet building with respect to  ``what'' the calculus does, and with respect to  ``how'' the spreadsheet generates specific values. ``Why'' the spreadsheet has been built with that specific shape, copes with a given layout choices, is intended to support a given business activity, or complex set of activities, and therefore implements a precise set of functionalities, etc., is completely embedded in the spreadsheet content itself, and current technologies do not offer solutions neither for the extraction of these information, when present, nor for formally coupling them with the actual editing functionalities.

For the sake of explanation, please consider the following naive scenario: Bob, stakeholder of a small sized corporation, develops a spreadsheet for calculating manufacturing cost estimates. According to his working experience, Bob has a general idea about how to build the calculus model, but he does not know exactly all the elements that actually take part into the calculus.  He arranges a spreadsheet and he tests it with input data coming from a pre-existing order. The test fails. As a consequence Bob manually changes some formula coefficients in order to obtain the desired value.  Afterwards he tests the same spreadsheet on other orders: it estimates correctly in some cases but not in some others. For fixing the problem, he modifies on the fly some spreadsheets formulas. The calculated cost estimation is now correct for the new order but not for the old one, so Bob needs to retrieve a copy of the old spreadsheet and compare the formulas with the new one in order to fix it. This time-consuming situation is very common in the spreadsheet development.

As future work we propose an evolutionary algorithm approach for helping the users in spreadsheets optimization. This approach is expected to be based on an interactive building of the fitness function. In what follows a general and qualitative overview of the idea is briefly sketched. Starting from the visual representation of the model, the user spotlights the values that are ``not correct'' (too high or too low) according to his mental model~\cite{johnsonlaird:hal-00702919}  - whatever ``mental'' is~\cite{cabitza2013leveraging} - of the problem. A population of fitting models is generated using an evolutionary algorithm that works on the formulas coefficients. The user can select some elements of the proposed models, and marks them as correct, or too high or too low. One of the models is selected according to the users interaction and a fitness function is built. The fitness function is used for the successive refinement of the model.

\begin{figure}
\centering
  \includegraphics[width=14cm]{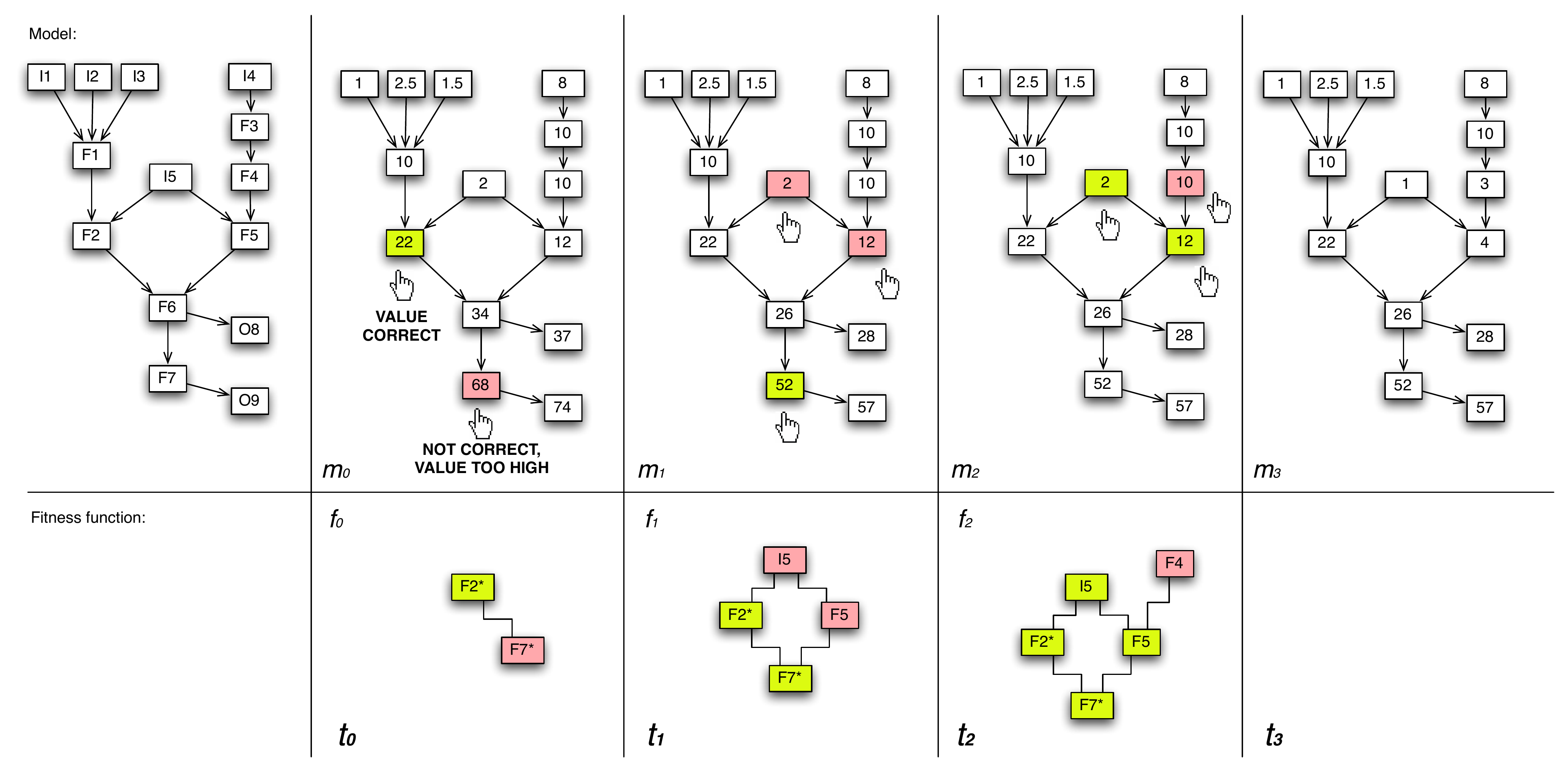}
   \caption{An example of spreadsheet evolution.}
   \label{fig:evo-example}
\end{figure}

An example is show in Figure~\ref{fig:evo-example}. At the beginning ($t_0$), the user selects two nodes of the graph $m_0$ and annotates one as correct and the other one as a too high value.  This annotations are used to create the fitness function $f_{0}$. Through the genetic algorithm a set of models are generated and one is randomly selected and proposed to the user. With another interaction, the users annotates that three nodes are correct and too high, so that a new fitness function $f_1$ is generated. This process continues until the user is satisfied by the obtained model.

%
%
\bibliographystyle{eptcs}
\bibliography{generic}
\end{document}